\PassOptionsToPackage{svgnames}{xcolor}
\documentclass[]{article}
\newtheorem{definition}{Definition}
\usepackage{dirtytalk}
\usepackage[hidelinks]{hyperref}
\usepackage{natbib}
\title{Sneak peek at the tig sequences:\\ useful sequences built from nucleic acid data.} 

\author{Camille Marchet$^{1\ast}$\\
{$^{1}$Univ. Lille, CNRS, Centrale Lille, UMR 9189 CRIStAL, F-59000 Lille, France}\\
{$^\ast$To whom correspondence should be addressed;}\\
{E-mail:  camille.marchet@univ-lille.fr.}
}
\date{}
\usepackage{tcolorbox}
\tcbuselibrary{skins,breakable}
\usetikzlibrary{shadings,shadows}

\newenvironment{myblock}[1]{%
\tcolorbox[colback=White,colframe=SandyBrown,%
title=#1]}%
{\endtcolorbox
}

\usepackage[moderate]{savetrees}
\begin{document}
\maketitle 


\begin{abstract}
This manuscript is a tutorial on tig sequences that emerged after the name "contig", and are of diverse purposes in sequence bioinformatics. We review these different sequences (unitigs, simplitigs, monotigs, omnitigs to cite a few), give intuitions of their construction and interest, and provide some examples of applications.
\end{abstract}

\section{Introduction}

The word \emph{contig} denotes contiguous sequences, a term which will be used later in sequence assembly, whose task is to reconstruct a genome from sequencing data.
It is first appearance of what I call a \emph{tig} sequence, from ~\cite{staden1980mew}, which quotes: \say{In order to make it easier to talk about data gained by the shotgun method of sequencing, we have invented the word 'contig'[...]}. 
Afterwards, other words with the tig suffix gradually appeared in the literature (although some of the sequences presented below do not have the tig suffix, but still fall in the scope). 
Certainly, these are related at least because they name a nucleotidic sequence with certain properties or guarantees with regard to computer science (for instance to represent a set) or to bioinformatics (for instance, to be linked to some biological properties).
They also share that their construction is based on computer science operations, whether it is on graphs representing sequences or using string properties.

This manuscript aims at presenting a unified vision on the collection of concepts and ideas related to tig sequences, that are currently used and under active research in sequence bioinformatics.
It gives the main concepts behind tig sequences, and proposes a categorization of these sequences based on their main purposes.
It also provides an intuition of the construction of tigs, and surveys their current use in bioinformatics along with some future directions.
For the sake of consistency and length, I chose to present in detail tigs related to $k$-mers and/or de Bruijn graphs (which covers the majority of tigs). Tigs defined on other structures are mentioned in the last section of the manuscript.

The target audience for this manuscript is mainly bioinformatics grad students, or scientists from connected fields who would like an introduction to the subject. They may or may not have a computer science background. While someone with a computed science background will also find interesting to seek details in the original paper, they can always come back to this tutorial which provides a quick and rather complete view of the tig landscape.
Computer science enthusiasts who may lack some concepts will appreciate that the manuscript remains not too technical and possibly makes concept more accessible than original papers. For any reader, tig sequences are a gateway to many major research themes in sequence bioinformatics, such as assembly, management of sequences collections and scalability, or algorithmic bioinformatics.

\section{Preliminary definitions}
In the following we formalize our framework and give some important definitions.
We work on finite strings (see Box 1.c) such as reads, genomic or transcriptomic sequences, genomes, over the genomic alphabet $\Sigma=\{A,C,G,T\}$ (if needed we replace $U$ by $T$ for convenience).
Tigs represent relevant nucleotidic units going beyond the smallest used substrings in indexing and assembly: $k$-mers. 

\begin{definition}{\bf $k$-mer.}
 A $k$-mer is a substring (see Box 1.c) of length $k$ from a given string of length $s \geq k$, i.e., $k$ consecutive nucleotides extracted from a position $p$ such that $0\leq p \leq s - k +1$.
\end{definition}

Many tigs were indeed created to deal with $k$-mer sets. An important concept related to $k$-mer sets is the de Bruijn graph (see Box 1 for fundamental graph notions). It is a well known object in the assembly field. The de Bruijn graph serves as a fundamental structure for the assembly of second generation sequencing data thanks to its efficiency in representing of $k$-mer sets. 

More generally, different graphs work on nucleotidic sequences. We will call \emph{sequence graphs} such graphs, which contain nucleotidic sequences in their nodes and link them in some way through edges. The rest of the presentation will be quite graph-oriented, but it is interesting to notice that ideas presented in the manuscript have twin concepts in stringology. We will mention a few of them.

Although there are several definitions of de Bruijn graphs in bioinformatics, here we propose one which is both well-accepted and useful to introduce other concepts of the paper.

\begin{figure}
\begin{myblock}{\textbf{Box 1. Graph \& string notions}}
\textbf{a. Graph.}
A graph is a couple of two sets $(\mathcal{V},\mathcal{E})$, $\mathcal{V}$ is a set of \emph{nodes} and a $\mathcal{E}$ is a set of \emph{edges}. An edge connects a sink node to a source node. A \emph{subgraph} is a selection of edges and nodes from a given graph. A \emph{directed graph} has orientation in its edges, represented by an arrow.

\textbf{b. Paths, Walks.}
We will use the term \emph{path} to denote a finite sequence of edges that joins a sequence of distinct nodes. We will call a \emph{walk} a similar sequence in which a node can be traversed several times. 

\textbf{c. Strings.}
A \emph{string} is an ordered sequence of characters over a given alphabet. For a string $S$ of size $l$, we will call $T$ a \emph{substring} of $S$ when $T$ is a string made of all $S$'s characters within two indices $0 \leq i \leq j < l$ kept in the same order than in $S$.
\end{myblock}
\end{figure}

\begin{definition}{\bf Node-centric de Bruijn graph (dBG) in bioinformatics.}
Given an input multiset of nucleotidic sequences $S$ on $\Sigma^{*}$, the de Bruijn graph is a directed graph $G_k(S) = (\mathcal{V},\mathcal{E})$ where $\mathcal{V}$ is a set of nodes and $\mathcal{E}$ a set of directed edges. $\mathcal{V}$ is the set of $k$-mers of $S$. For $x,y\in \mathcal{V}$, an edge $(x, y)\in E$ if and only if $x[2,k]=y[1,k-1]$. In the following, we will simply call such structure a de Bruijn graph.
\end{definition}

\begin{figure}[ht]
    \centering
    \includegraphics[width=1\textwidth]{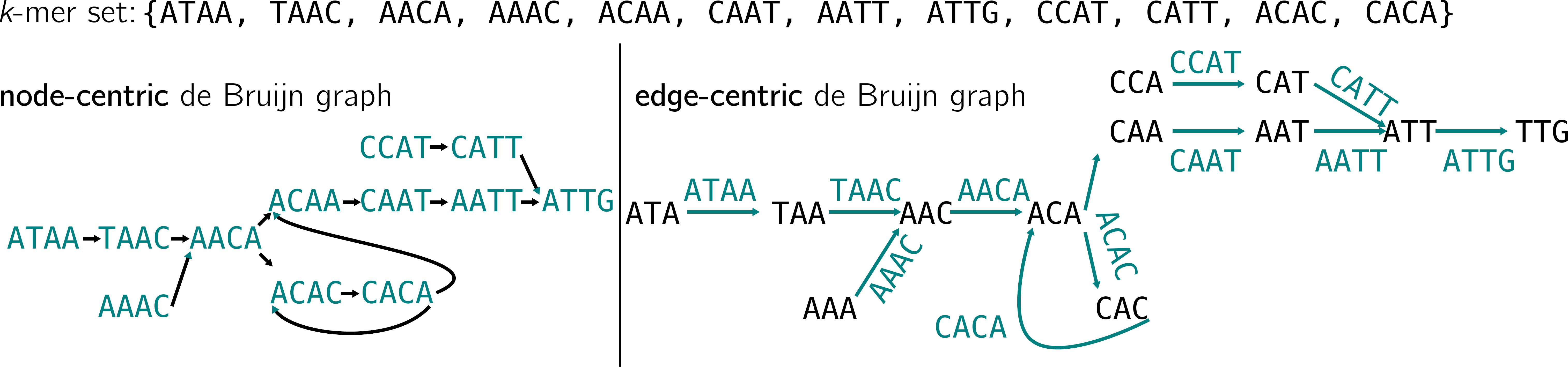}
    \caption{A $k$-mer set (top), the corresponding de Bruijn graphs ($k=4$, $k$-mers shown in blue). Nodes contain strings, oriented edges are represented using arrows that can be labeled. We show a a node-centric representation (left, $k$-mers are nodes and overlaps are implicit in edges) and an edge-centric representation (right, $k$-mers are labels of edges).}
    \label{fig:dbg1}
\end{figure}

In our definition, a de Bruijn graph built upon a list of sequences $S$ contains all distinct $k$-mers of $S$ as nodes (called node-centric definition after~\cite{chikhi2021data}, see Figure~\ref{fig:dbg1}). Importantly, node-centric means that the set of edges is implicit as can be inferred from the nodes. Indeed, nodes are connected if the $k-1$ suffix of a source node $x$ matches exactly the $k-1$ prefix of a sink node $y$. 
Some works on de Bruijn graph use the edge-centric definition were $k$-mers are the labels of edges (see Figure~\ref{fig:dbg1} for an example).

The de Bruijn graph in bioinformatics is a subgraph of the de Bruijn graph as generally defined in computer science. Indeed, instead of showing all possible words on the given alphabet, it presents only $k$-mers from the original data. Therefore in the node-centric definition, it is subgraph built from a selection on the nodes of the general de Bruijn graph.

In practice, for instance in the case of $k$-mers extracted from reads, we do not know the original strand of the $k$-mer (forward or reverse). De Bruijn graphs can account for this information by being  bidirected, a property that allows to encode all possible encountered overlaps (forward-forward, forward-reverse, reverse-forward, reverse-reverse) between nodes.
In this manuscript, for the sake of simplicity, we avoid on purpose to present the full case which takes into account forward and reverse sequences. In practice some application keep only \emph{canonical $k$-mers}, i.e. the smallest lexicographic version between a $k$-mer and its reverse complement (for instance, for AGGT and its reverse complement ACCT, the canonical $k$-mer will be ACCT).

\section{Introduction to tigs with unitigs}

\subsection{Link between tigs and sequence graphs}
An early assembly paper (\cite{myers2000whole}) introduces the first tig of interest for us, unitigs, as “[c]ollections of fragments whose arrangement is uncontested by overlaps from other fragments”. 
Assembly relies on graph objects, \emph{assembly graphs}, which are built from reads and connect their nodes (reads, called fragments in Myers' quote, or pieces of reads) to reconstruct larger genomic pieces. Assembly graphs rely on overlaps between reads, the difficulty lying in repeated regions that can yield several overlaps for a given read. 
We will see that tig sequences are very related to sequence graphs, notably to the de Bruijn graph presented earlier, as these graph offer a framework to build tigs.

\subsection{Unitigs}\label{sss:unitigs}
\begin{figure}[ht]
    \centering
    \includegraphics[width=1\textwidth]{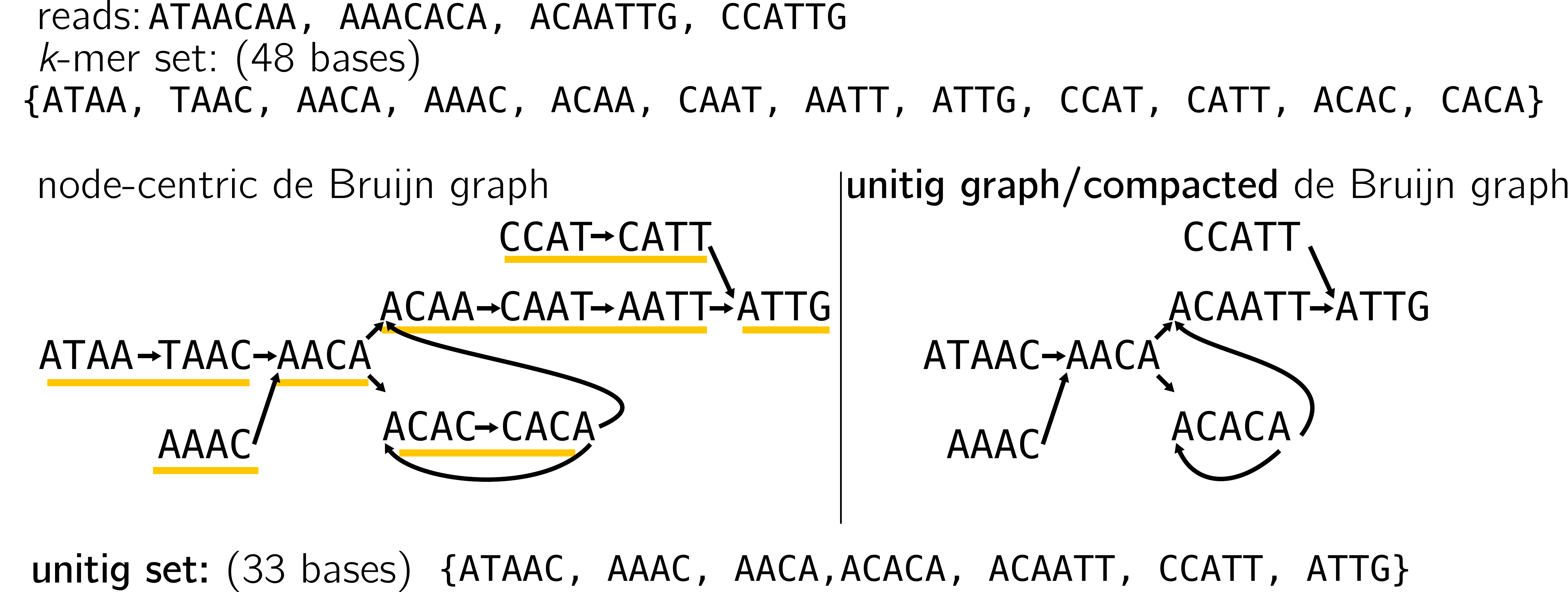}
    \caption{A read set and the extracted $k$-mer set (top), the corresponding de Bruijn graph ($k=4$, left) similar to Figure~\ref{fig:dbg1}, paths leading to unitigs are shown in yellow in the graph, the unitig graph (right) and the unitig set (bottom). Each unitig starts or ends either at graph dead ends or when a bifurcation occurs, i.e. a node has several sinks or sources. See how the graph creates connections between reads (for instance the end of the first one becomes connected to the beginning of the third one). Note that for unitig ACACA, a second equivalent compaction is possible: CACAC, starting by $k$-mer CACA then compacting using the final "C" of ACAC.}
    \label{fig:dbg}
\end{figure}

\underline{\textbf{Unitigs}} are built from a de Bruijn graph. In assembly, unitigs are usually considered as "safe" sequences, because it seemed that one can assemble their $k$-mers without ambiguity (we will see later in section~\ref{ss:omnitigs} that this notion has been nuanced). Put another way, there should be only one way to complete a sequence using $k$-mers from inside a unitig.
When an ambiguity (a bifurcation in the graph, see examples in Figure~\ref{fig:dbg}) happens, the unitig is stopped and other ones start. 
These sequences are often output during the inner steps of an assembler, before being further elongated into contigs, which usually involve more heuristic choices in their construction (however, in other assembly schemes, contigs can also arise from different sequences than unitigs). 
\\

A path (see Box 1.a for a reminder on paths) in a de Bruijn graph is a sequence of nodes joined by edges, such that the edges in the sequence are all directed the same way, and each node appears only once. Unitigs are maximal simple paths in the de Bruijn graph, simply put, the longest possible sequence of nodes that can be traversed by following edges, stopping at (and including) any node which has more than one incoming edge or more than one outgoing edge.

We call \emph{compaction} the operation on $k$-mers which yields unitigs, whose resulting nucleotides are written in a single, larger or equal to $k$, string.

First we present the glue operation. It builds strings larger or equal to $k$ from a sequence of consecutive $k$-mers by adding only nucleotides which bring novel information to the string (as compared to redundant nucleotides in $k$-mer overlaps).
\begin{definition}{\bf Glue operation.}\label{def:glue}
Given a sequence of nodes $n_0, \ldots n_i$ of a de Bruijn graph, the glue operation creates a glue node $n_g$ containing a string $s$, which copies $n_0$'s $k$-mer. Then it consecutively concatenates each last nucleotide of $n_1, \ldots n_i$  to $s$.
\end{definition}

Then we define a maximal simple path on which the glue operation will be applied.
\begin{definition}{\bf Maximal simple path.}
Let $G$ be a directed graph. A maximal simple path in $G$ is a sequence of nodes $u=n_0, \ldots n_p \in G$ such that $n_0$ (respectively $n_p$) has more than one in-going edge or more than one outgoing edge. Any node $n_i$ of $u$ such that $1 \leq i \leq p-1$ is entered and left only one time.
\end{definition}

The compaction is the operation which glues nodes from maximal simple paths in de Bruijn graphs. The created unitig spells the same string than the created glue node.
\begin{definition}{\bf Compaction.}\label{def:compact}
Let $\mathcal{G}$ be a de Bruijn graph.
Let $u$ be a maximal simple path in $\mathcal{G}$.
A compaction creates a glue node $n_g$ by applying a glue operation to the nodes sequence of $u$. 
\end{definition}

A unitig graph, or compacted de Bruijn graph, is a graph built from compaction operations. The $k$-mers sequence of maximal simple paths of de Bruijn graphs are then \emph{compacted} into unitigs.
\begin{definition}{\bf Unitig graph or compacted de Bruijn graph.}
A compacted de Bruijn graph $G$ is a directed graph built from a de Bruijn graph $\mathcal{G}$ by computing the unitig set from $\mathcal{G}$. $G$'s nodes are unitigs (after compaction).
$G$'s edges represent $k-1$ suffix/prefix overlaps between $G$'s nodes.
\end{definition}
 Therefore, the de Bruijn graph can be converted to a graph of unitigs and the inverse operation is possible as well. Constructing a graph of unitigs from a de Bruijn graph can have multiple solutions (unitig sets, see an example in Figure~\ref{fig:dbg}).

The creation of a compacted de Bruijn graph can be seen as a graph simplification of a de Bruijn graph, as first formalized in~\cite{medvedev2008ab}. Then, \cite{cazaux2014indexing} applies these simplifications to de Bruijn graphs (the paper calls them \emph{contracted} de Bruijn graphs) and \cite{chikhi2016compacting} proposed the first efficient implementation for constructing these graphs. 

An example of compaction in Figure~\ref{fig:dbg} starts from $k$-mer ATAA, and adding \say{C} and \say{A} from the two next $k$-mers of the unitig to obtain ATAACA (as seen in the definition, we use the term unitigs for maximal unitigs).

\subsection{Example of tig property: efficient \texorpdfstring{$k$}--mer set representation}

With unitigs, we have seen how a tig sequence can be extracted from a set of $k$-mers. Aside from assembly, unitigs have been used as computational objects to handle $k$-mer sets. 
Indeed, from a $k$-mer set $S$, a unitig set $U$ can be built, and all $k$-mers of $S$ can be retrieved once and only once in $U$. $U$ is said to represent the set of $S$. Representing exactly sets of $k$-mers is currently a major use of tigs since it allows precise computational management for these sequences. 

\begin{definition}{\bf Spectrum-preserving string set (SPSS).}
Given an input string set containing $k$-mers $S$, a spectrum preserving string set of $S$ is a plain text representation (e.g. a set of strings) that has the same set of $k$-mers as $S$, and does not contain duplicate $k$-mers. 
\end{definition}
The SPSS notion has been introduced recently~\cite{rahman2020representation} and has gained traction since then, as some research field have been looking into representing more efficiently sets of $k$-mers. Most SPSS do not handle multiplicity. Thus, they preserve the set of $k$-mers from a list of nucleotidic sequences, but not the $k$-mer multiset. There are exceptions that are presented in the following.
\\
The most obvious SPSS is the $k$-mer set itself, and we noticed that a set of unitigs from a de Bruijn graph is a SPSS as well. 
At worst (in a very fragmented graph), they use as much nucleotides to represent the $k$-mer set as the $k$-mer set itself, but usually, they represent it in a more compacted and efficient way (in Figure~\ref{fig:dbg} we used 33 nucleotides in comparison to the 48 of the $k$-mer set).
The representation is not optimal since nucleotides from the overlaps are represented several times.
For the sake of simplicity, Figure~\ref{fig:dbg} presents very small $k$-mers, but the burden of redundancy increases with real-life-sized $k$-mers (usually in the 21-51 nucleotides range for second generation sequencing reads).

In order to discuss the next tig, notice the red substrings in Figure~\ref{fig:dbg} , that show some redundancy that remains in the representation. Such redundancy occurs because unitigs still share a $k-1$ overlap on their extremities.

\section{Tig sequences for \texorpdfstring{$k$}--mer sets computational management}

\subsection{Simplitigs and UST: nearly optimal SPSS}\label{ss:simpli}
Keeping up with the idea of SPSS, and of representing the $k$-mer set, different works showed that there are better objects than unitigs to minimize the number of nucleotides in the representation. 
Two works, for \underline{\textbf{simplitigs}} (\cite{brinda2016novel, bvrinda2021simplitigs}) and for \underline{\textbf{UST}} (after Unitig-STitch, \cite{rahman2020representation}) described a solution simultaneously, though independently. UST and simplitigs are therefore used for space footprint reduction when storing $k$-mer sets, and provide source code\footnote{UST: \url{https://github.com/medvedevgroup/UST/blob/master/README.md}, simplitigs: \url{https://github.com/prophyle/prophasm}}.

Strings larger than $k$-mers can be found by following paths in the graph.
Then, finding a SPSS is equivalent to finding a set of paths such that it covers all nodes of the de Bruijn graph (called a \emph{path cover}), and then realize a compaction of the nodes in each path. With the constraint of each $k$-mer appearing only once in the final string set, this means that a $k$-mer should be used only once in the whole path cover (therefore called \emph{distinct;} path cover). 
A simple, non minimal solution is that each path starts and ends in a single node. This would correspond to the $k$-mer set itself being a SPSS. The unitig set is another distinct path cover.
The UST/simplitigs intuition is that unitigs can themselves be compacted to obtain longer sequences and reduce the number of $k-1$ redundancies. 

\begin{figure}[ht]
    \centering
    \includegraphics[width=1\textwidth]{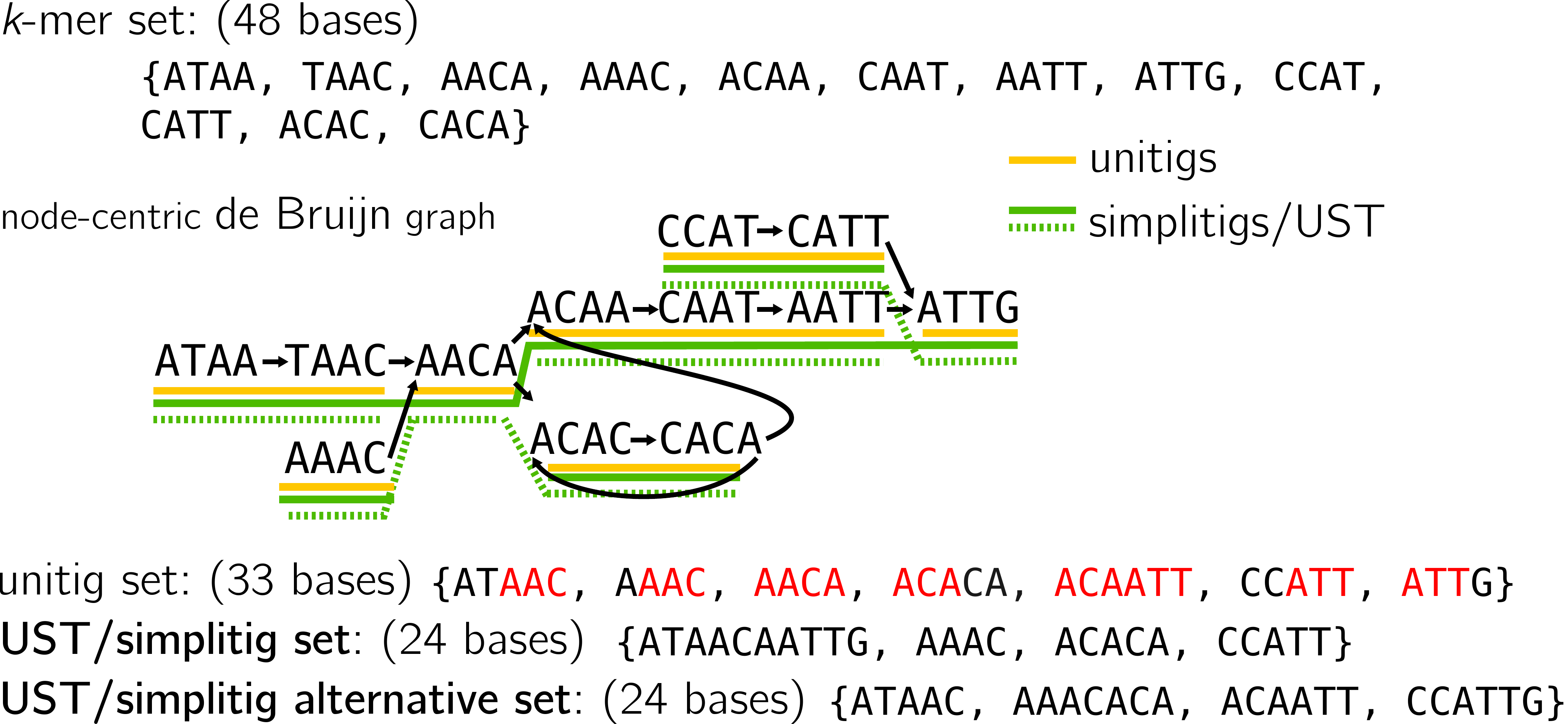}
    \caption{A $k$-mer set (top), and the simplitigs/UST built from the corresponding de Bruijn graph ($k=4$, similar to Figures~\ref{fig:dbg1} and~\ref{fig:dbg}). Two possible compaction schemes are shown (green/dotted green). Unitigs are shown for the comparison (yellow). Red parts of the unitig set correspond to redundant nucleotides that belong to $k-1$ overlaps between unitigs. Redundant parts are reduced in simplitigs/UST.}
    \label{fig:simplitig}
\end{figure}

Both papers propose a greedy algorithm to achieve that compaction. 
It means that for a given node, the algorithm looks for an edge leading to a sink node, explores this node and uses it for compaction if it has not been explored before. The compaction is continue until there is no more node to be explored, and each node is visited only once.
Interestingly these works also show that the greedy method is close to the lower bound (the lower bound being the theoretical minimum number of compacted strings). 
Figure~\ref{fig:simplitig} presents an example of these sequences compared to unitigs.

\subsection{Eulertigs and matchigs: minimal SPSS and other formulation of the problem}

\subsubsection{Minimal SPSS}
This section is more arduous than the previous. In a first paragraph, we intend to provide a simple intuition and let the reader decide whether they want to follow up to the technicalities.

Works following simplitigs/UST aim at minimizing the number of characters to encode the set of $k$-mers, with SPSS properties. To minimize the SPSS, we need the distinct path cover to be have the smallest cardinality, i.e. to include as few paths as possible. Intuitively, fewer paths lead to more $k$-mer compaction, hence less redundancy. Eulertigs give boundaries and a linear algorithm (see Box 2) for the minimal SPSS problem. To date, UST/simplitigs implementations are a bit faster than Eulertigs\footnote{Eulertigs implementation: https://github.com/algbio/matchtigs} Results show that all these methods compute SPSS in less than 10 minutes on the $k$-mers of thousands of \textit{E. coli} genomes~(\cite{schmidt2022eulertigs}).
\\

\begin{figure}[ht]
    \centering
    \includegraphics[width=1\textwidth]{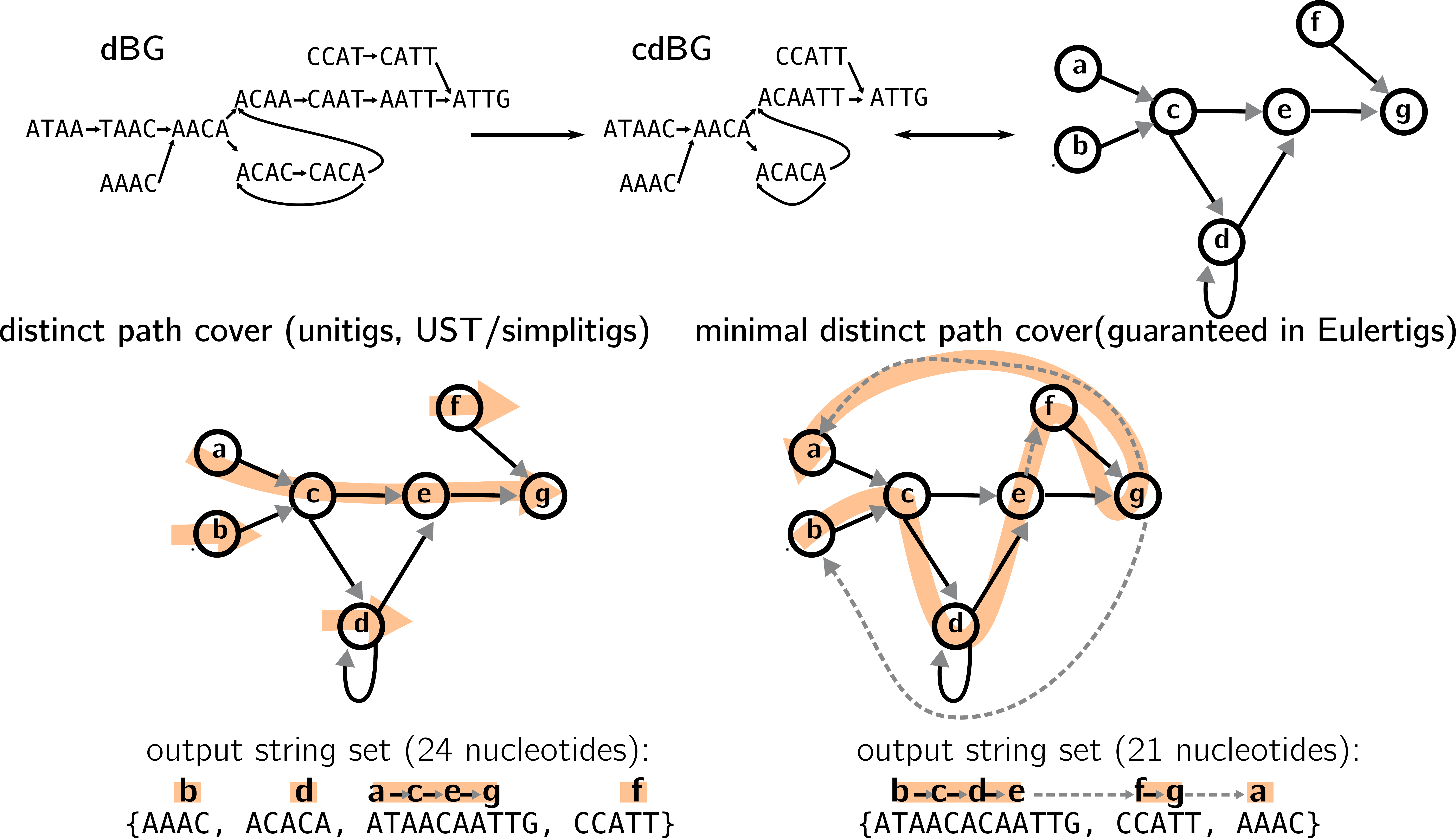}
    \caption{At the top, we show the correspondence between the graph notation we will use and the initial de Bruijn graph (similar to Figures~\ref{fig:dbg1},\ref{fig:dbg} and \ref{fig:simplitig}). The de Bruijn graph is represented as an unitig graph (middle), and each unitig is given a letter to simplify the representation. At the bottom, we compare a distinct path cover that could have been computed with simplitig/UST solution for instance (same than in Figure~\ref{fig:path}, top right), with a minimal distinct path cover given by Eulertigs. Keep in mind that for the sake of simplicity we show an example on a node-centric representation, but for Eulertigs the graph has actually to be converted to an edge-centric de Bruijn graph. The Eulertig solution adds special edges (dotted lines) to balance the graph: each node has as many in-going edges than out-going edges (the paper shows it is always possible to draw all needed edges). A cycle in orange can go through any edge, special or not. Then, the cycle is broken into a set of paths by removing special edges. The result is a minimal distinct path cover.}
    \label{fig:euler}
\end{figure}

I first give an informal intuition of the minimal distinct path cover problem problem solving. In order to find such a set, some graph traversal can guarantee each needed property, i.e. having a cover on the distinct $k$-mers, and yielding the least possible strings.
One solution is to modify the graph by adding necessary special edges so that we can find a cycle that goes through all nodes of the graph once (see Figure~\ref{fig:euler}). That would be called a hamiltonian cycle. Some constraints would be necessary, as the special edges should be added only to nodes which have not yet reached a balance between in-going and out-going edges. Then, by removing the special edges of the cycle we obtain a set of paths which is necessary minimal (see bottom right of Figure~\ref{fig:euler}).
Luckily, in the special case of a de Bruijn graph defined in bioinformatics on the nucleotide alphabet, this problem can be solved linearly.
In the general case, finding hamiltonian cycles on a graph is NP-complete which means that it is very unlikely to find an algorithm solving this problem in a reasonable amount of time.

\begin{figure}[ht]
    \centering
    \includegraphics[width=1\textwidth]{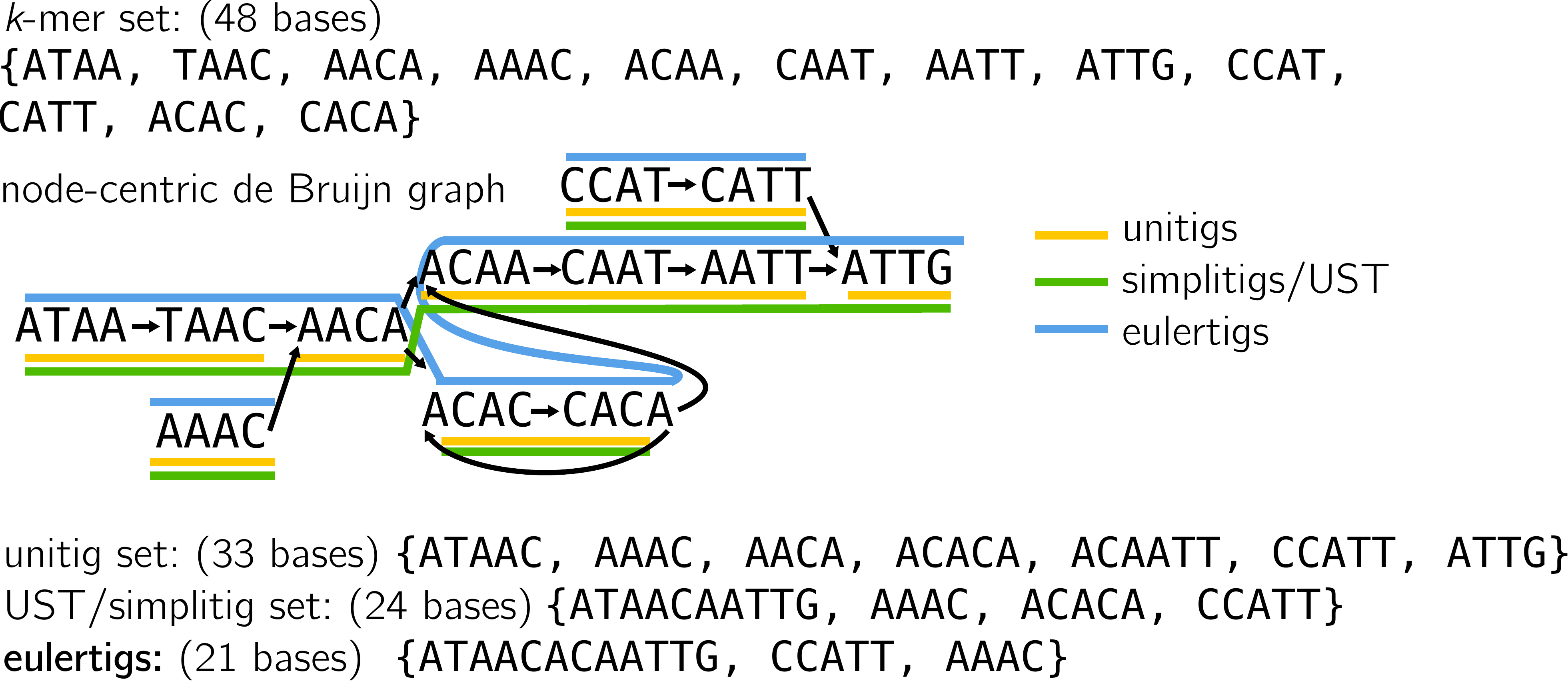}
    \caption{Comparison between Eulertigs (blue), simplitigs/UST (green) and unitigs (yellow) on a de Bruijn graph (graph is similar to Figures~\ref{fig:dbg1},\ref{fig:dbg} and \ref{fig:simplitig}).}
    \label{fig:path}
\end{figure}

In practice, solving the problem means switching from the node-centric de Bruijn graph to an edge-centric de Bruijn graph (from a given node-centric dBG built from $(k+1)$-mers of the data, build a dBG where all possible $k$-mers are in edges), and solving the same problem on edges (eulerian cycle).
\underline{\textbf{Eulertigs}} paper~(\cite{schmidt2022eulertigs}) takes advantage of this property to propose an algorithm to find a minimal SPSS on a $k$-mer set time-linear in the size of the SPSS.\\

This correspondence between eulerian and hamiltonian graphs in the special case of de Bruijn graphs of $k$-mers has been a source of confusion on how assembly works~(\cite{medvedev2021eulerian}).
The difference here is that the goal is not to assemble genomes but indeed to find a good way to represent successive $k$-mers, regardless of the biological meaning of the output string.\\

Here, I share an observation which is not mentioned in the literature to my knowledge. Two earlier works had produced all necessary material to solve the minimal SPSS problem, as done by Eulertigs. \cite{crochemore2010algorithms} defined an equivalent stringology problem to the problem of finding a distinct path cover of smallest cardinality: finding \emph{shortest common superstrings on a set of words}. It is defined on 2-mers and solves the problem using eulerian cycles. Then \cite{golovnev2013approximating} showed how to solve this problem generally on $k$-mers with $k-1$ overlaps. Eulertigs can be seen as inheriting from these works, and applying the solution to close the question as defined in the SPSS context.

\begin{figure}
\begin{myblock}{\textbf{Box 2. Asymptotic notations.}}

{We say an algorithm or a method solves a problem in \textbf{linear time} (or $\mathcal{O}(n)$) in the size of an object $n$ when in the worst case of running the algorithm, the time cost function of the algorithm grows more slowly than a linear function in infinity, up to a multiplicative constant.

Informally, we can say that the time needed is proportional to the size of $n$ at worst. $n$ is a parameter, it can be the size of the input or something else. Typically, reading a vector until a given element is found or the end of the vector is reached takes a time proportional to $n$ the size of the vector at worst.\\}

There are other asymptotic behaviors.
 Practically in bioinformatics, we aim at asymptotic behaviours close to $\mathcal{O}(n)$ or below (e.g. $\mathcal{O}(log(n))$ or running in constant time) at least for large objects, for performance matters.

\end{myblock}
\end{figure}

\subsubsection{Other \texorpdfstring{$k$}--mer sets}
A previous work (\underline{\textbf{matchtigs}} in~\cite{schmidt2021matchtigs}) proposed to relax the distinct path cover property, by finding a path cover which allows a $k$-mer to be re-used in different paths of the set (the $k$-mer should still appear once in a given path). The intuition is that some $k$-mers act as bridges that reduce the redundancy. Therefore matchtigs do not fall in the definition of SPSS we gave, but still represent the set of $k$-mers.

\begin{figure}[ht]
    \centering
    \includegraphics[width=1\textwidth]{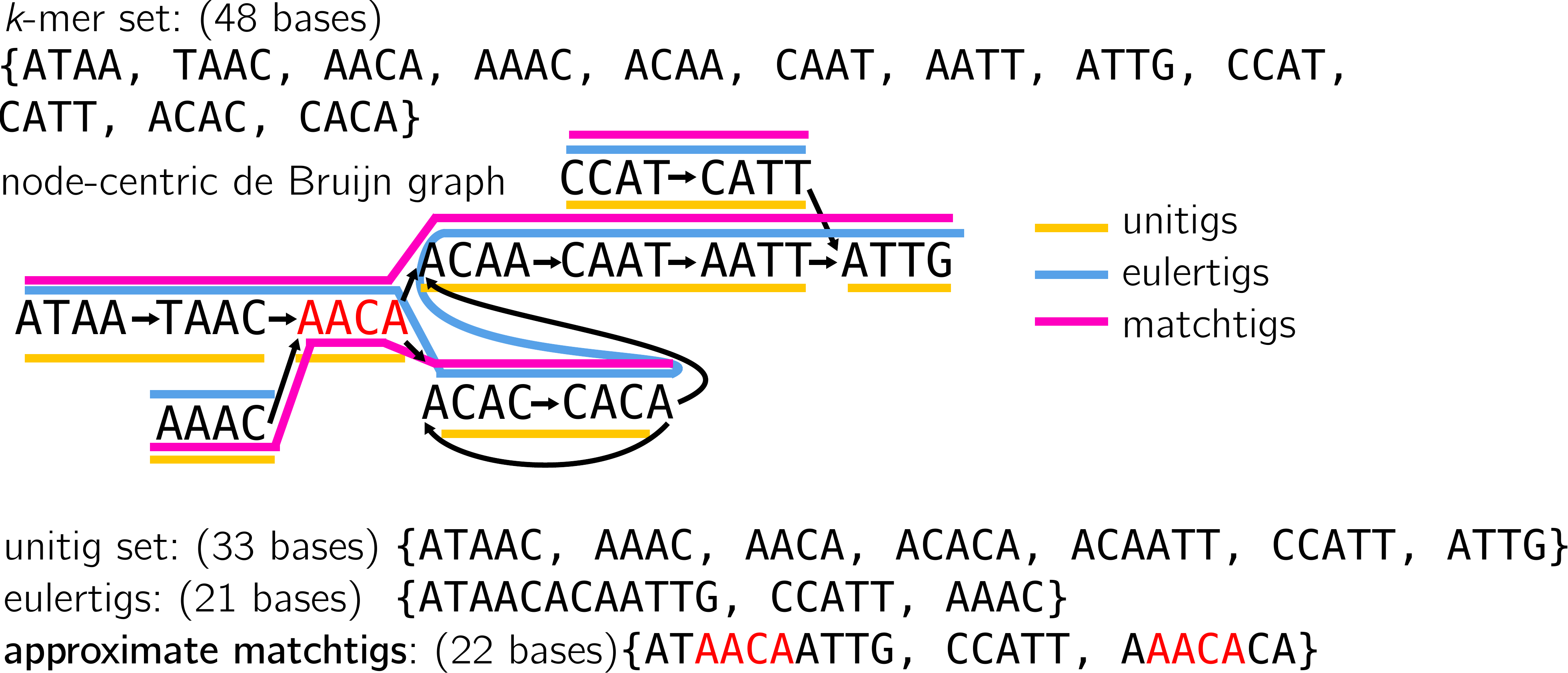}
    \caption{
    Comparison between matchtigs (purple), Eulertigs (blue), and unitigs (yellow) on a de Bruijn graph (graph is similar to Figures~\ref{fig:dbg1}, \ref{fig:dbg}, \ref{fig:simplitig} and \ref{fig:path}). In the purple path cover: notice that the red $k$-mer AACA is included in two different paths. Considering that AACA is used already in the path spelling ATAACACAATTG after compaction, it prevents to split the path set after the $k$-mer AAAC and before ACAC in comparison to unitigs, therefore nucleotides are saved in these $k$-mers overlaps.}
    \label{fig:matchtig}
\end{figure}

The idea behind the construction is similar to Eulertigs, finding eulerian cycles in the graph. Matchtigs are actually an earlier solution than Eulertigs, using the relaxed property of using a $k$-mer several times.
More precisely, a given $k$-mer can be used in several paths of the cover.
Eulertigs then showed that this condition was not necessary to obtain minimal representations.
The paper proposes minimal matchtigs and approximate matchtigs, the latter do not guarantee a minimal representation but yet represent an improvement over UST/simplitigs. See an example of approximate matchtigs in Figure~\ref{fig:matchtig}, which shows the main difference with other approaches: $k$-mer redundancy in the set representation.

\subsection{Tig sequences for parallel computation: super--\texorpdfstring{$k$}--mers}

For performance purposes, one can need to perform operations on $k$-mers in a parallel way. In this case, it is interesting to have an efficient method to dispatch $k$-mers in balanced buckets and then to easily retrieve each $k$-mer's bucket.
\underline{\textbf{Super-$k$-mers of unitigs}} are substring from unitigs built for this purpose.
Super-$k$-mers are built by compacting all consecutive $k$-mers of a unitig that share a similar \emph{minimizer}. Super-$k$-mers of unitigs are often less efficient than unitigs in terms of nucleotide minimization to represent the set of $k$-mers, since it is not their main goal.
Historically, the first super-$k$-mers to be introduced are the super-$k$-mers from reads~(\cite{li2015mspkmercounter}). They differ from the super-$k$-mers of unitigs since they are built from the read sequences.
Observe that super-$k$-mers from unitigs are also a SPSS.
In order to associate these SPSS to one and only one partition or bucket, \emph{minimizers} are used.

\begin{definition}{\bf Minimizer.}
Given $k$, $m \leq k$ and a function $h$ defining an order, a minimizer is the smallest $m$-mer with respect to the order given by $h$, that appears within a $k$-mer when screening positions $0\ldots k-m+1$.
\end{definition}
Minimizers have been introduced in two independent contributions~(\cite{roberts2004reducing,schleimer2003winnowing}). 
We must stress that there exist other and more general definitions and use cases for minimizers than the one presented here, notably where minimizers are used to sample $k$-mers (for instance in sequence comparison and mapping methods). 
In the examples of the manuscript, $h$ defines the lexicographical order. However, in practice, $h$ is often a random hash function, which has been shown to have better properties for minimizers~(\cite{schleimer2003winnowing}) (therefore, $m$-mers are mapped to integers and the smallest integer is selected).

\begin{figure}[ht]
    \centering
    \includegraphics[width=0.8\textwidth]{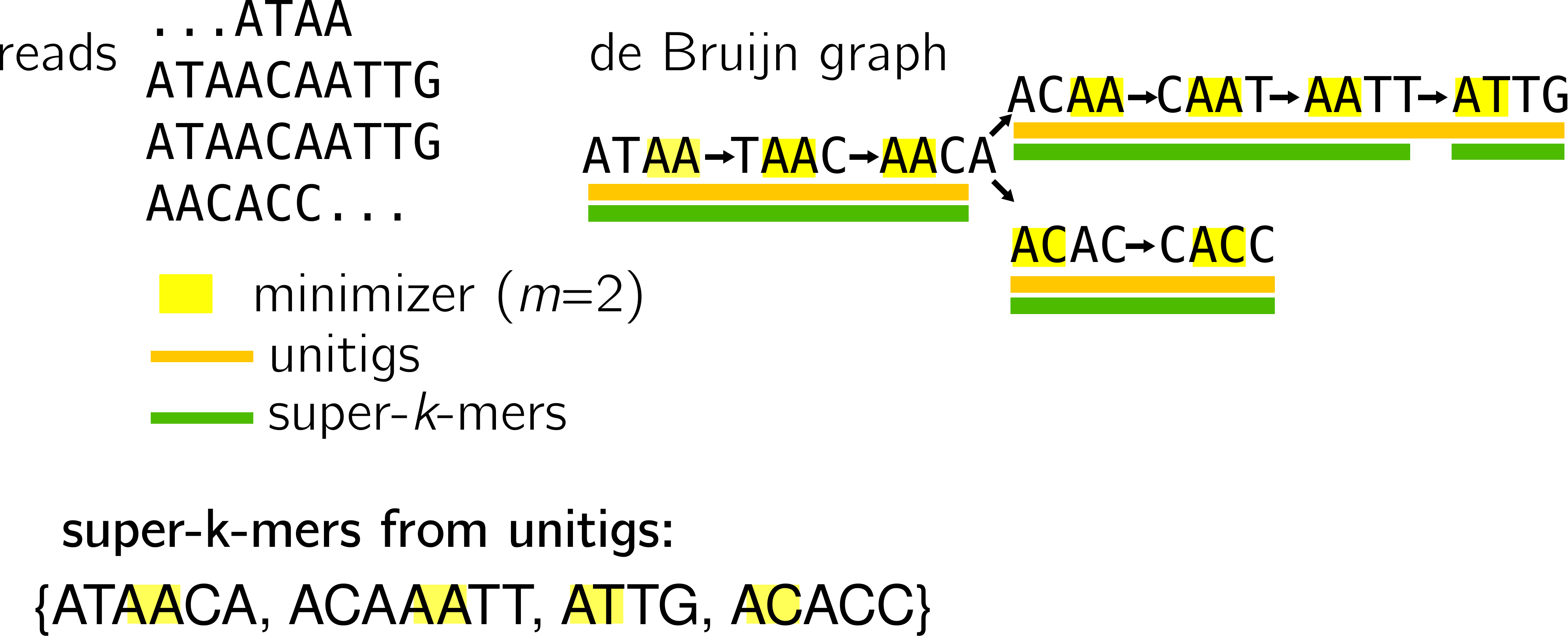}
    \caption{From a read set, $k$-mers are extracted and a de Bruijn graph can be built. Super-$k$-mers from unitigs are shown at the bottom and in green paths. Minimizers ($m=2$) are highlighted in yellow, we use the lexicographical order for this example. By following the green paths, we can see that super-$k$-mers break for two reasons: end of a unitig or switch of minimizer. The yielded super-$k$-mers of this example could be stored in three buckets labeled "AA" (ATAACA and ACAAATT), "AT" (ATTG) and "AC" (ACACC).}
    \label{fig:superk}
\end{figure}

Thus, with a wisely chosen minimizer scheme, one can dispatch $k$-mers in balanced buckets per minimizer. An example of super-$k$-mers is provided in Figure~\ref{fig:superk}.

\section{Tig sequences in biological sequence analysis}
In the previous section we reviewed helpful tigs for computational management of $k$-mer sets. In this section, we focus on tigs whose motivation is to bring valuable information to sequence analysis, often -but not only- in the context of assembly.
We can start by noting this section and the previous ones share overlaps.
De Bruijn graphs have been introduced in bioinformatics in the context of short-read assembly, and therefore unitigs are tightly linked to assembly as well. \underline{\textbf{Contigs}} are more loosely defined as the set of strings in the final output of an assembler. \underline{\textbf{Scaffolds}}, which are contigs aggregates forming longer sequences, have been called occasionally \underline{\textbf{supercontigs}} (see for instance~\cite{jaffe2003whole}). We will present other tigs that were introduced more recently, such as monotigs and omnitigs.

\subsection{Tig sequences with additional biological information: monotigs}
We now consider more information and assume that $k$-mers can come from different samples that are all pooled in a graph.
We can consider the $k$-mer sample presence/absence profiles in the graph, i.e., whether a $k$-mer is present a in given dataset or not. The datasets can be ordered and listed, and this information can be encoded using bit vectors. For instance, a $k$-mer presence/absence profile can be encoded as 0110, which would mean there are 4 datasets, and this $k$-mer is absent from the first and fourth, and present in the second and third.
\underline{\textbf{Monotigs}} were introduced in~\cite{marchet2020reindeer} in order to create a SPSS that also guarantees that all $k$-mers in a string of the SPSS have the same bit vector profile.
A side effect is that they can record multisets or sets of sets of $k$-mers.

\begin{figure}[ht]
    \centering
    \includegraphics[width=1\textwidth]{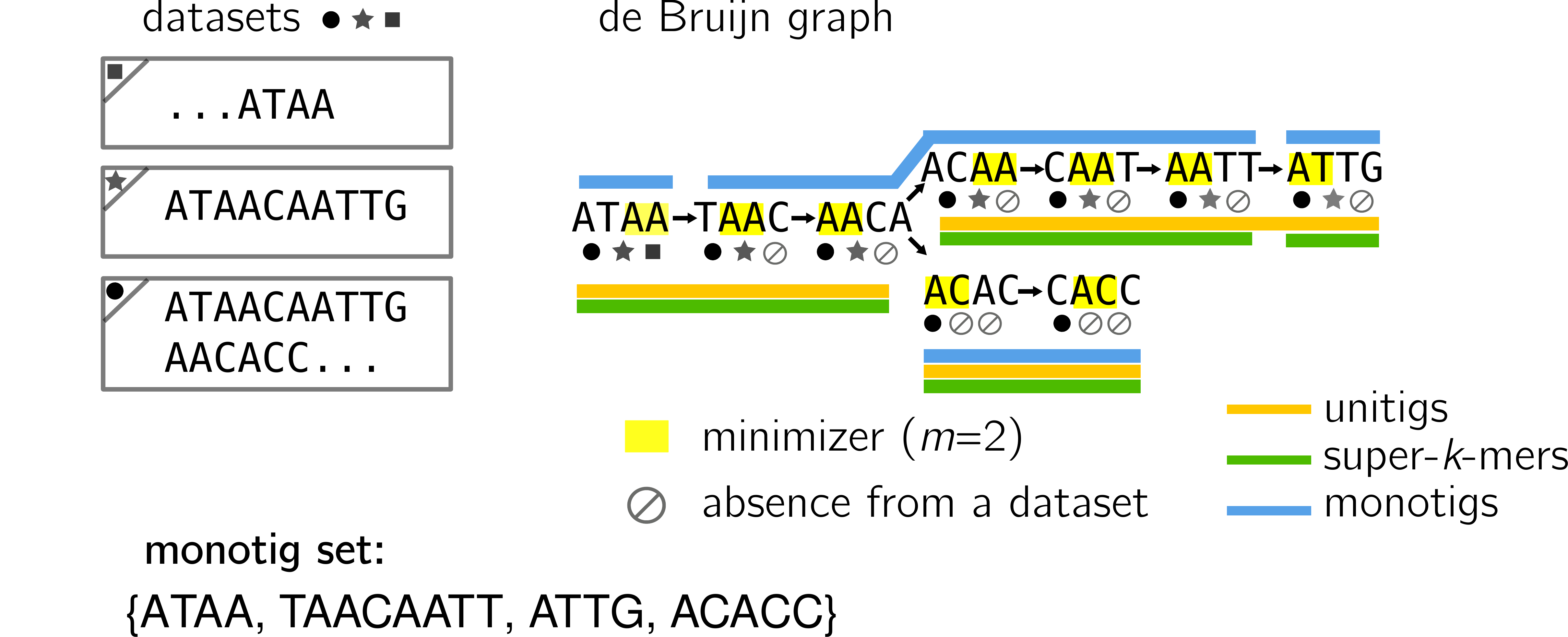}
    \caption{Three different datasets (samples, genomes, \ldots) are represented using a symbol (circle, star, square). A de Bruijn graph ($k=4$, similar to Figure~\ref{fig:superk}) is built by pooling $k$-mers from all samples. Monotigs are built using minimizers ($m=2$), and here, the presence/absence patterns are taken into account. In reality, monotig can also handle abundances. We can see that monotigs can span several unitigs, and break whether because a minimizer changes or because the profile changes.}
    \label{fig:mono}
\end{figure}

Monotigs are showed in Figure~\ref{fig:mono}.
We notice that unitigs built from several datasets can contain $k$-mers that have different presence/absence profiles, for instance yielded by chimeric sequences in the assembly graph. For instance in Figure~\ref{fig:mono} the leftmost unitig ATAACA contains $k$-mers present in all three datasets and $k$-mers not present in the square dataset.
This phenomenon happens as $k$-mers from separate samples share $k-1$ overlaps. Monotigs fix this discrepancy by storing two different information for these $k$-mers.

\subsection{Omnitigs and macrotigs: genomic assembly in de Bruijn graphs}\label{ss:omnitigs}
We leave the SPSS realm in this section, but we continue to review the tig sequences.
Here we make the choice to spend more time on omnitigs as they are defined on an already defined object, de Bruijn graphs. Tigs on other assembly graphs are briefly described in section~\ref{ss:direction}.

These tigs' motivation is to represent a "safe" set of sequences, i.e., that will be found in any assembly solution from a de Bruijn graph. 
Indeed, assemblers add different heuristics on top on their formal definitions of assembly based on sequence graphs in order for their operations to be run in a reasonable amount of time. Therefore, it is interesting to be able to extract some sequences that can show guarantees despite the possibly empiric choices implemented in assemblers.
But recent work also shows that strings that were considered "safe", such as unitigs, are in fact not in all cases, since they can be substrings absent from the initial genome~(\cite{rahman2022assembler}). 
Previous works already demonstrated empirically that some overlaps between $k$-mers were spurious and had implemented correction steps guided by reads to remove them from assemblies (\cite{bankevich2012spades}).

In short, when compacting unitigs in so-called contigs, the assembler has to make choices at ambiguous bifurcations.
\underline{\textbf{Omnitigs}} will be found in any contig set that is a solution of an assembly graph, regardless of the compaction choices.

\subsubsection{Omnitigs definition}
\begin{figure}[ht]
    \centering
    \includegraphics[width=0.8\textwidth]{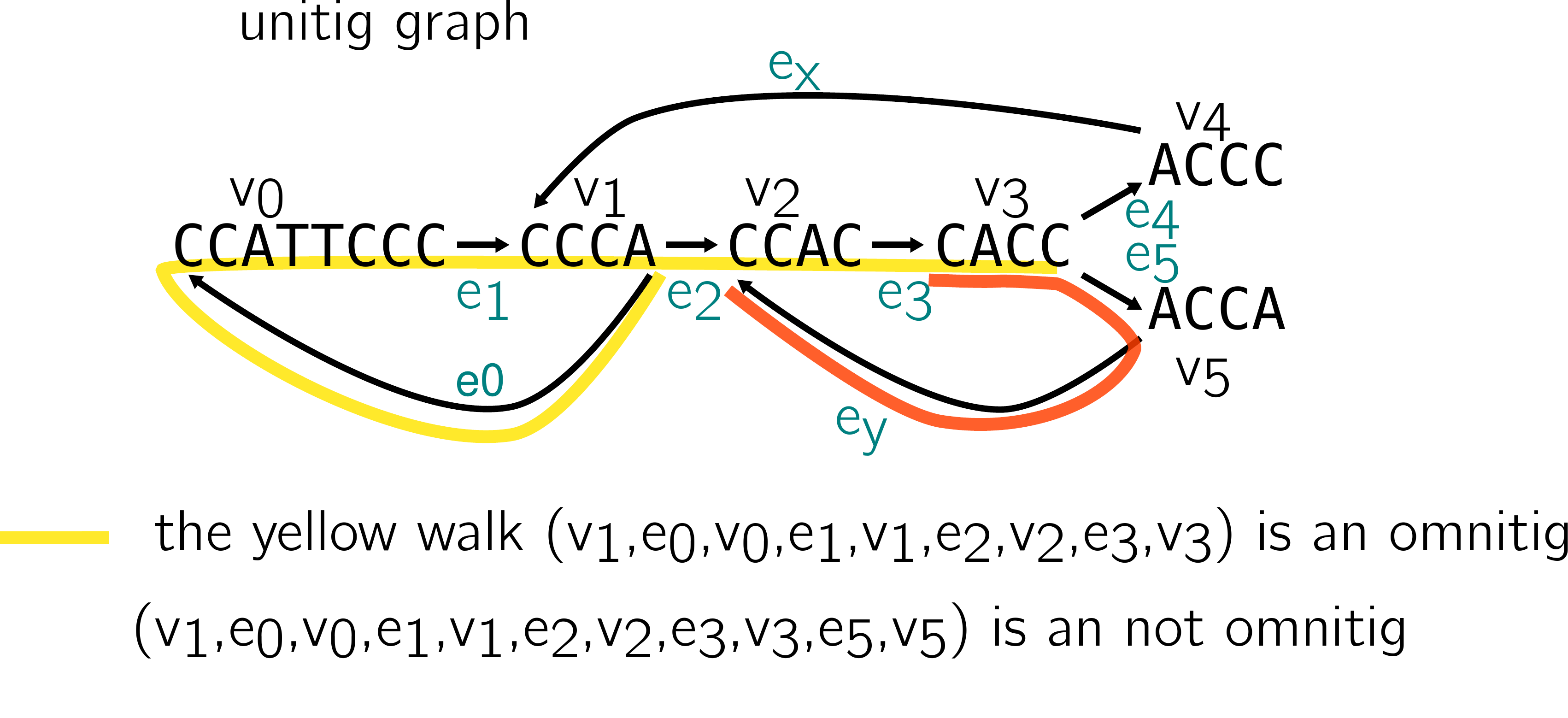}
    \caption{We work on a unitig graph built from a de Bruijn graph (see subsection~\ref{sss:unitigs}). The yellow walk shows an omnitig. ($v_1, e_0, v_0, e_1, v_1, e_2, v_2, e_3, v_3, e_5, v_5$) is not an omnitig, indeed with this node sequence, there exist a path from v3 to v2 such that e2 is not included: ($v_3, e_5, v_5, e_y, v_2$) (red path).}
    \label{fig:omni}
\end{figure}

In a unitig graph, omnitigs (in their edge-centric definition) are a walk from node $n_0$ to node $n_{w-1}$ (with an edge $e_l = (n_{l-1}, n_l), \: 0\leq l \leq w-1$), such that for all $1 \leq i \leq j \leq w-1$, 
 there is no path that allows to go from $v_j$ to $v_i$ without having $e_{j+1}$ as first edge, and $e_i$ as last edge. 
 To illustrate this, we use two examples inspired from ~\cite{tomescu2017safe, cairo2017optimal} in Figure~\ref{fig:omni}, one shows a walk that is an omnitig, the second shows a walk that is not an omnitig.

To describe the longest "safe" sequences from contigs, one can compute the set of maximal omnitigs.

\subsubsection{Introduction to omnitig construction}

To introduce omnitigs construction, we will present the Y to V operation which is a way compact the graph. This operations happens to be at the core of constructing omnitigs, although in some cases it is not sufficient, then more technicalities can be found in omnitigs papers. We will show Y to V operation's idea only.\\

\begin{figure}[ht]
    \centering
    \includegraphics[width=1\textwidth]{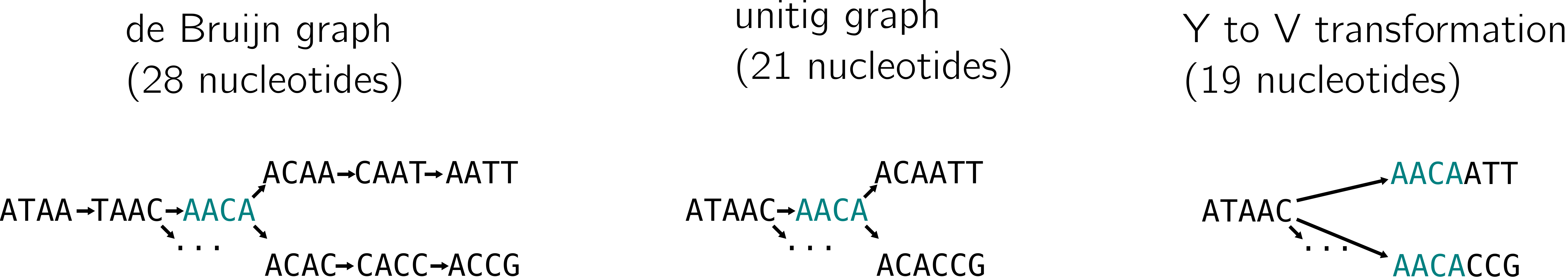}
    \caption{Y to V transformation. We start from a de Bruijn graph, converted to a unitig graph. In this branching pattern, the common source for the different sink nodes is the node AACA in blue. Note that in the unitig graph, all nucleotides of this node are redundant because of unitig overlaps. This node is duplicated and compacted to the sinks to create two distinct sink nodes in Y to V. transformation. This transformation consumes less nucleotides because the redundant overlaps AAC and ACA appear fewer times.}
    \label{fig:yv}
\end{figure}

Let's focus on bifurcations in de Bruijn graphs. In simple cases as shown in Figure~\ref{fig:yv}, the two sink nodes have a single, unambiguous source. 
Y to V operation proposes to duplicate the content of this parent node and to compact it with the children, then remove the parent.

\begin{figure}[ht]
    \centering
    \includegraphics[width=1\textwidth]{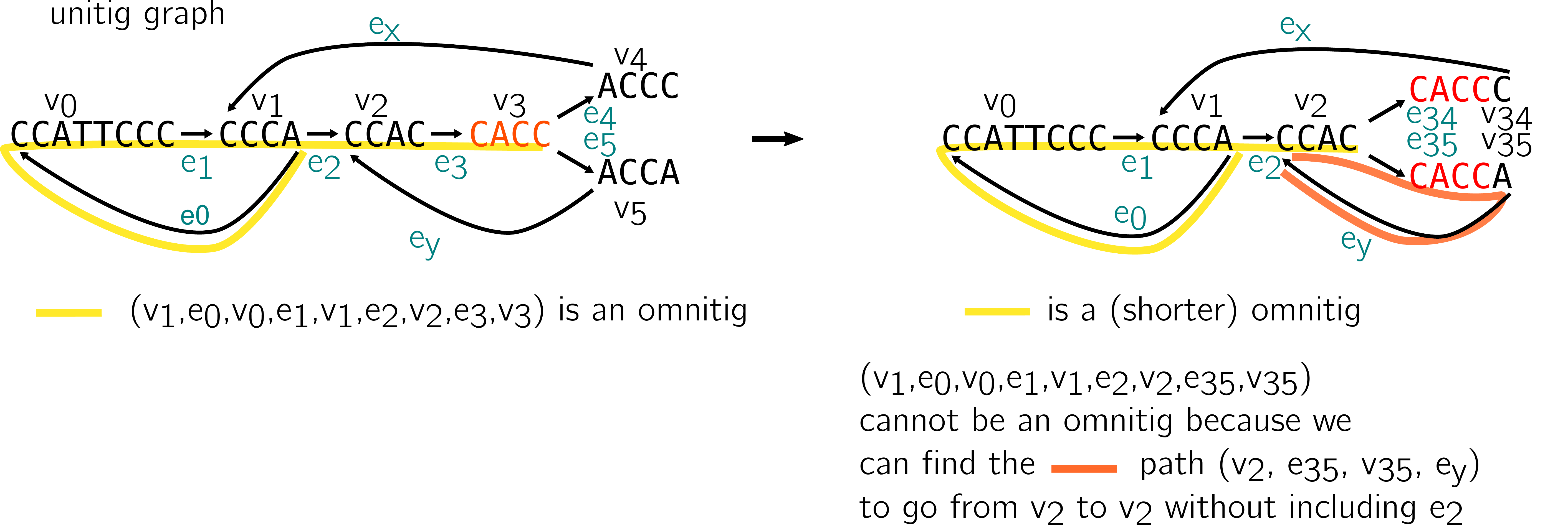}
    \caption{A unitig graph on the left undergoes a Y to V transformation. 
($v_1, e_0, v_0, e_1, v_1, e_2, v_2, e_{35}, v_{35}$)  cannot be an omnitig because we can find the orange path ($v_2, e_{35}, v_{35}, e_y$) going from $v_2$ to $v_2$ without including $e_2$.}
    \label{fig:omni2}
\end{figure}

In Figure~\ref{fig:yv}'s example, the blue node is duplicated in the two children nodes. 
AACAATT and AACAACG are two "safe" sequences that will be found in contigs.
Y to V operation do almost all the work in building omnitigs, however an extra step is sometimes necessary.
That is because Y to V is not always optimal as it can prevent finding the longest "safe" sequences in some cases. In Figure~\ref{fig:omni2} we show an illustration of why Y to V operation is sometimes not enough, and can prevent from finding the maximal omnitigs in a graph.

Several papers which followed on omnitigs~(\cite{obscura2018safe,diaz2018assembling,cairo2019optimal}), and \underline{\textbf{macrotigs}} were recently introduced~(\cite{cairo2020genome}) as a way to compute maximal omnitigs in $\mathcal{O}(n)$ time, with $n$ the number of edges in the graph.

\section{Applications of tigs in (meta-)genomics and transcriptomics}
\subsection{Use of tig sequences in the current literature}
\subparagraph*{Compact representation for $k$-mer set data-structures.}
In associative data-structures, pairs of $(key,values)$ can imply  an explicit storage of the keys. These is the case for efficient dictionaries that emerged as highly space and time saving data-structures for $k$-mers. 
Since these techniques cannot handle duplicate input keys, they are complemented with efficient key ($k$-mer) set representations. SPSS, with their set properties, are interesting to that extent as they permit to record $k$-mers in a compact way, and to access them using quick operations on bits. Therefore the UST/simplitig concept has been adopted in recent $k$-mer dictionaries (\cite{10.1093/bioinformatics/btab217,pibiri2022sparse,pibiri2022weighted}). Being very recent and in practice marginally space-saving compared to the previous solution, Eulertigs are not yet integrated in such implementations.

A \textit{k-mer file format} (KFF)\footnote{\url{https://github.com/Kmer-File-Format/}} introduces a new step in $k$-mer management. Relying on UST and super-$k$-mers, it has been proposed by~\cite{dufresne2022k} and is currently integrated in several $k$-mer tools for compression, $k$-mer counting and pre-processing. 
It improves on the computation time and practicality of previous SPSS implementations. While most space efficient SPSS require around 3 bits per $k$-mer on a human genome, KFF takes 11-17 bits per $k$-mer and down to $\sim$7 bits when gz compressed, and is readily usable to encode them in binary.

\subparagraph*{Compression.}
General  compression schemes like gzip cannot be optimal to compress $k$-mer sets since they do not exploit the specific nucleotide redundancy in these sets. 
To improve compression, UST were used for $k$-mer sets~(\cite{rahman2021disk})\footnote{implementation: \url{https://github.com/medvedevgroup/ESSCompress}} and another paper~(\cite{kitaya_et_al:LIPIcs.WABI.2021.12}) introduces its own SPSS construction for compression.

\subparagraph*{$K$-mer partitioning.}
Super-$k$-mers are a powerful tool to allow efficient partitioning and parallel algorithms on sequence data. They are used in~\cite{10.1093/bioinformatics/btab217} to accelerate the construction of a $k-$mer dictionary. Moreover, since all $k$-mers of a super-$k$-mer can be treated in a single query, they also allow a speed-up when looking for sequences in these data-structures. The metagenomic classifier based on $k$-mers, Kraken~(\cite{wood2014kraken}) uses the same principal for fast $k$-mer retrieval. This way of partitioning $k$-mers becomes more and more frequent~(\cite{10.1093/bioinformatics/btab156}).

\subparagraph*{De Bruijn graph representation.}
Noticing that more efficient $k$-mer set representation directly leads to better de Bruijn graph representations, tigs have been integrated in graph compaction methods~(\cite{khan2021scalable}). Unitigs have been used in short-read assembly context for a long time~(\cite{chikhi2016compacting}). 
Globally, de Bruijn graph representation is a field in itself, with supplementary features and needs compared to plain SPSS representation (for instance, operations to navigate the graph are needed). Some methods are reviewed in~\cite{chikhi2021data}.

\subparagraph{Assembly.}
De Bruijn graph and related objects have been a major asset in assembly, especially for short reads. 
Disjointigs are used in long read assembly from Pacific Biosciences and Oxford Nanopore technologies in several assemblers~(\cite{kolmogorov2019assembly, kolmogorov2020metaflye, bankevich2021lja}). With haplotigs, we witness the progress of assembly since with some instances of long read data (for instance HiFi reads), currently being able to phase haplotypes~(\cite{cheng2021haplotype}).

\subparagraph{Collections of biological sequences.}
Colored de Bruijn graphs can be informally defined as de Bruijn graphs built from more than one sample or genome, which $k$-mers are labelled with their dataset of origin. Monotigs are used in colored de Bruijn graphs~(\cite{marchet2020reindeer, marchet2021data}) as inner objects guaranteeing $k$-mer features (counts, presence/absence patterns) across different datasets. Other SPSS have been implemented in colored de Bruijn graphs (for instance \cite{schmidt2021matchtigs}).

\begin{figure}[ht]
    \centering
    \includegraphics[width=1\textwidth]{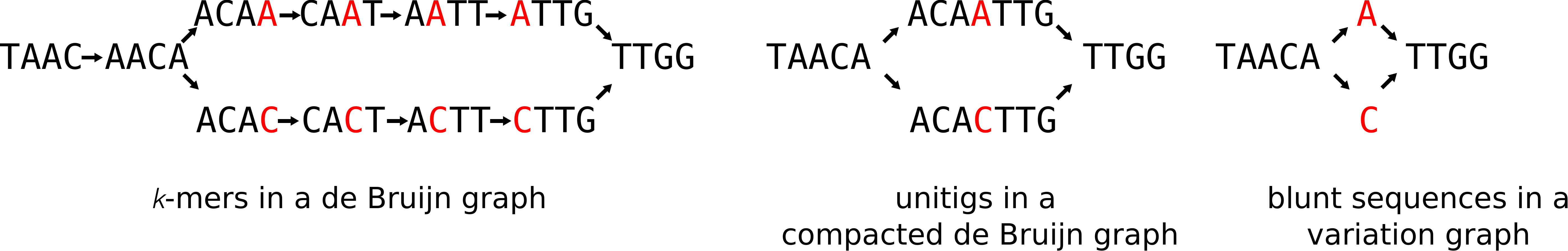}
    \caption{Difference between de Bruijn graph (left), compacted de Bruijn graph (middle) and variation graph with blunt sequences (right). An A/C mutation is shown in red.}
    \label{fig:blunt}
\end{figure}

Pangenome graphs are another object which represents collections of sequences. One major difference between colored de Bruijn graphs and pangenome graphs (variation graphs and others) comes from the fact that the input of pangenomic graphs is ordered and supposedly free from sequencing errors (a list of reconstructed genomes, with chromosome coordinates), while the data input to de Bruijn graphs is not (an unordered set of reads, sometimes contigs).
Some structures make the bridge between the two, such as de Bruijn graphs built from reference genomes (\cite{khan2021cuttlefish} for the most recent). For computational needs, it can be interesting to move from one representation to the other. 
Principally, assembly graphs represent sequences in their nodes and overlaps in edges. Overlaps are important in this type of graphs because they materialize the support reads give to the adjacency of two sequences. 
On the contrary, pangenomics graph usually work from references and use edges to show direct adjacency between genomic subsequences, without overlaps.
Such sequences are called blunt sequences (see Figure~\ref{fig:blunt}), with recent efforts to transition from one to the other (\cite{eizenga2021walk}).

\subsection{Possible directions}\label{ss:direction}
\subsubsection{Other algorithmic questions for the SPSS}
The SPSS application may seem closed because Eulertigs brought a time-linear solution to optimally solving the minimal SPSS problem, yielding a set of strings representing all $k$-mers once in the smallest cardinality. However, matchtigs showed that the problem can be defined differently, being changing the rules (e.g., recording a multiset of $k$-mers). For instance, all proposed solution work separately on each connected component (subgraphs that are not connected by any edge) of the de Bruijn graph, but it could be done differently by adding extra edges between them. Efficient indexing of the different tigs is also vastly open.

SPSS focused on the base-wise efficiency aspect. However, there can be other applications to SPSS, for instance there could be other ways to order the $k$-mers regarding assembly, so that overlaps or repeats are easier to detect. As an example, a recent unpublished preprint (\cite{Diaz2022.09.06.506758}) defines contigs on variable order de Bruijn graphs (i.e., with nodes of variable $k$ length).

\subsubsection{Tigs beyond \texorpdfstring{$k$}--mers and de Bruijn graph}
First of all, we must mention that some reviewed tigs can be defined on other graphs than the de Bruijn graph. It is the case for omnitigs and matchtigs, which can be computed on another type of assembly graph, the string graph. The reader might have noticed that SPSS applications were mainly directed towards short-reads, second generation sequencing data for which the volume is a bottleneck. Future directions include anticipating similar questions with very large $k$-mers (more than 100 nucleotides) or objects more error-tolerant than $k$-mers (such as strobemers \cite{sahlin2021strobemers}).  

With short reads, the de Bruijn graph has been the major assembly graph structure in assembly. However, recently, with the advent of third generation sequencing long reads, assembly graphs include novel and more diverse sequence graphs. Therefore, other tigs have been defined or brought back in the spotlights. For instance, the overlap graph is a graph used in long read assembly, on which \underline{\textbf{disjointigs}} (\cite{kolmogorov2019assembly}) are built. In their construction they look like simplitigs and UST. However, their purpose is different as they help with assembly. They are used to solve repeats by finding in which order repeated regions should be traversed.
\underline{\textbf{Haplotigs}} are contigs whose inner variants come from the same haplotype. Although the definition is quite old~(\cite{makoff2007detailed}), the concept has been more frequently used since long reads assembly took over. Indeed, haplotypes are more prone to be solved with long sequences covering series of variants. Overall, we can expect seeing more tigs associated with long reads assembly in the future.

Finally, defining tigs as relevant biological units is another exploratory purpose. For instance in RNA-seq, we need strong experimental validation to assert whether monotigs or other close objects can help studying genes by splitting them in different functional modules.

\section*{Acknowledgments}
This article followed active discussions on Twitter and a related blog post. Many thanks to Rayan Chikhi for his feedback and suggestions on the initial blog post. I also would like to thank Bastien Cazaux and Antoine Limasset for their valuable inputs on the manuscript, and Jamshed Kahn for pointing out the first contig mention in a paper on Twitter. Finally, a huge merci to Paul Medvedev for suggesting me to write this piece and for providing important feedback.
\nocite*{}
\bibliographystyle{plainnat}
\bibliography{main}
\end{document}